\documentstyle[preprint,aps,12pt]{revtex}
\tightenlines
\begin{document}
\draft

\preprint{YITP-98-37,\\hep-ph/9807378}

\title{Non-factorizable contributions \\
in hadronic weak decays of charm mesons}
 
\author{ K. Terasaki}

\address{Yukawa Institute for Theoretical Physics, \\
Kyoto University, Kyoto 606-01, Japan}
\date{July 1998}
\maketitle

\begin{abstract} 
Two body decays of charm mesons are studied by describing their 
amplitude in terms of a sum of factorizable and non-factorizable
ones. The former is estimated by using a naive factorization while 
the latter is calculated by using a hard pseudo-scalar-meson 
approximation. The hard pseudo-scalar-meson amplitude is given by a 
sum of the so-called equal-time commutator term and surface term 
which contains all possible pole contributions of various mesons, not 
only the ordinary $\{q\bar q\}$ but also four-quark 
$\{qq\bar q\bar q\}$, hybrid $\{q\bar qg\}$ and glue-balls. 

Naively factorized amplitudes for the spectator decays which lead to 
too big rates can interfere destructively with exotic meson pole 
amplitudes and the total amplitudes can reproduce their observed 
rates. The non-factorizable contributions can supply sufficiently
large contributions to the color suppressed decays which are strongly 
suppressed in the naive factorization. A possible solution to the 
long standing puzzle that the ratio of decay rates for 
$D^0\rightarrow K^+K^-$ to $D^0\rightarrow \pi^+\pi^-$ is around 2.5 
is given by different contributions of exotic meson poles. 

\end{abstract}
 
\vskip 0.5cm
\newpage

\section{Introduction}

Nonleptonic weak decays of charm mesons have been studied extensively 
by using the so-called factorization (or vacuum insertion) 
prescription\cite{BSW,NRSX}. However, recent semi-phenomenological 
analyses\cite{CLEO,Kamal} in two-body decays of $B$ mesons within the 
framework of the factorization suggest that the value of $a_2$ to 
reproduce the observed branching ratios for these
decays\cite{CLEO,PDG} should be larger by about a factor 2 than the 
one with the leading order (LO) QCD corrections\cite{BSW,NRSX,HW} 
where the color degree of freedom $N_c = 3$ and that its sign should 
be opposite to the one in the large $N_c$ limit although the 
phenomenological value of $a_1$ is very close to the one expected in 
the same approximation. [$a_1$ and $a_2$ are the coefficients of four 
quark operators in the effective weak Hamiltonian in the 
Bauer-Stech-Wirbel (BSW) scheme\cite{BSW,NRSX} which will be reviewed 
briefly in the next section.] The above fact implies that the large 
$N_c$ argument fails, at least, in hadronic weak decays of $B$
mesons. Since the large $N_c$ argument is independent of flavors, it 
also does not work in nonleptonic weak decays of charm mesons. 
Therefore dominance of the factorized amplitude in charm decays 
looses its theoretical support since, in charm meson decays, the 
large $N_c$ argument is the only one known theoretical background to 
support the factorization prescription\cite{Large-N}. In fact, a 
naive application of the factorization to charm decay amplitudes 
causes many problems, for example, too strong suppression for the 
color mismatched decays like $D^0 \rightarrow \pi^0\bar K$ and 
$D^0 \rightarrow \pi^0\pi^0$, too big rates for decays described by 
the so-called spectator diagrams like the Cabibbo-angle favored 
$c \rightarrow s\,+\,(\bar du)_1$ 
and Cabibbo-angle suppressed 
$c \rightarrow d\,+\,(\bar du)_1$ and 
$c \rightarrow s\,+\,(\bar su)_1$, 
too small ratio (less than unity) of the rates 
$\Gamma(D^0\rightarrow K^+K^-)$ to 
$\Gamma(D^0\rightarrow \pi^+\pi^-)$ 
although the observed value is around 2.5, etc., where $(\bar q'q)_1$
denotes a color singlet pair of $\bar q'$ and $q$. To get rid of 
these problems, the factorization has been implemented by multiplying 
the factorized amplitudes by phase factors arising from final state
interactions. However, the final state interactions are realized by 
dynamical contributions of various hadron states. 

Using a hard pseudo-scalar-meson approximation, the present author
has studied dynamical contributions of various hadron states to charm 
meson decays and has given a hint to solve the above problems in 
charm meson decays\cite{TBD,QTD}. However, in these analyses, the 
amplitudes did not include the factorizable contributions so that the 
results were not necessarily satisfactory, {\it i.e.}, these analyses 
could not satisfactorily provide an overall fit to the observed rates 
of Cabibbo-angle favored and suppressed decays of charm mesons. 
On the other hand, a recent analysis in hadronic weak decays of $B$ 
mesons\cite{BDP} by assuming that their amplitude can be given by a
sum of factorizable and non-factorizable ones suggests that, only in 
some processes under a particular kinematical condition (a heavy 
quark goes to another heavy quark plus a pair of light quark and 
anti-quark with sufficiently high energies like 
$\bar B\rightarrow D\pi$ and $D^*\pi$), the factorization works 
well while non-factorizable contributions are important in the other
decays, in particular, in color suppressed decays. It seems to imply 
that the naive factorization is not guaranteed by the large $N_c$
arguments but it works well under some special kinematical
condition\cite{Dugan}. In this article, we reanalyze two body decays 
of charm mesons describing their amplitude by a sum of factorizable
and non-factorizable ones. We will review briefly the naive 
factorization and list the factorized amplitudes for two body decays 
of charm mesons in the next section. Non-factorizable amplitudes in a 
hard pseudo-scalar-meson approximation will be presented in section 
III. In section IV, we will compare our result with experiments. 
A brief summary will be given in the final section. 

\section{Factorized amplitudes}

Our starting point in this article is to describe the two body decay 
amplitude by a sum of factorizable and non-factorizable
ones\cite{BDP}, 
\begin{equation}
M_{\rm total} = M_{\rm fact} + M_{\rm non-f}.          \label{eq:AMP}
\end{equation}
The factorizable amplitude $M_{\rm fact}$ is evaluated by using 
the factorization in the BSW scheme\cite{BSW,NRSX} in which the 
relevant part of the effective weak Hamiltonian responsible for the
charm decays is given by 
\begin{equation}
H_w^{BSW} 
= {G_F\over\sqrt{2}}
\Bigl\{a_1O_1^{(s'c)H} + a_2O_2^{(s'c)H} 
                                + (penguin\,\,term) + h.c.\Bigr\}.     
                                              \label{eq:BSW}
\end{equation}
It can be obtained by applying the Fierz reordering to the 
usual effective Hamiltonian, 
\begin{equation}
H_w
= {G_F\over\sqrt{2}}
\Bigl\{c_1O_1^{(s'c)} + c_2O_2^{(s'c)} 
                                + (penguin\,\,term) + h.c.\Bigr\}, 
                                              \label{eq:WH}
\end{equation}
where $c_1$ and $c_2$ are the Wilson coefficients of the normal
ordered four quark operators,   
\begin{equation}
O_1^{(s'c)} = :(\bar ud')_{V-A}(\bar s'c)_{V-A}:,  \qquad
O_2^{(s'c)} = :(\bar s'd')_{V-A}(\bar uc)_{V-A}:. 
                                             \label{eq:FQO}
\end{equation}
Here $d'$ and $s'$ denote weak eigen states of the down and strange
quarks, respectively, and 
$(\bar ud)_{V-A}=\bar u\gamma_\mu (1 - \gamma_5)d$, etc. 
We do not explicitly show the possible penguin term through which 
$s$-channel pole contributions of a scalar glue-ball can play a role 
in Cabibbo-angle suppressed decays, since its explicit expression 
will not be needed. The quark bilinears in $O_1^{(s'c)H}$ and 
$O_2^{(s'c)H}$ are treated as interpolating fields for the mesons and 
therefore should be no longer Fierz reordered. The coefficients $a_1$ 
and $a_2$ in Eq.(\ref{eq:BSW}) are given by 
\begin{equation}
a_1 = c_1 + {c_2 \over N_c}, \qquad a_2 = c_2 + {c_1 \over N_c}.
                                                     \label{eq:WC}
\end{equation}
The LO QCD corrections lead to $a_1\simeq 1.09$ and $a_2\simeq 0.09$ 
for $N_c=3$\cite{NRSX}. 

The factorization prescription in the BSW scheme leads to the
following factorized amplitude, for example, for the 
$D^+(p) \rightarrow \bar K^0(p')\pi^+(q)$ decay, 
\begin{eqnarray}
&& M_{\rm fact}(D^+(p) \rightarrow \bar K^0(p')\pi^+(q))   \nonumber\\
&&\qquad = {G_F \over \sqrt{2}}U_{cs}U_{ud}
\Bigl\{ 
a_1\langle \pi^+(q)|(\bar ud)_{V-A}|0\rangle 
\langle \bar K^0(p')|(\bar sc)_{V-A}|D^+(p) \rangle         \nonumber\\
&&\hspace{3cm} + a_2\langle \bar K^0(p')|(\bar sd)_{V-A}|0\rangle
\langle \pi^+(q)|(\bar uc)_{V-A}|D^+(p) \rangle
\Bigr\}.
                                                    \label{eq:FACT}
\end{eqnarray}
where $U_{ij}$ is the Cabibbo-Kobayashi-Maskawa (CKM) matrix 
element\cite{CKM} which is taken to be real in this article since 
$CP$ invariance is always assumed. Factorizable amplitudes for the 
other two-body decays also can be calculated in the same way. 
 
To evaluate the factorized amplitudes, we use the following
parameterization of matrix elements of currents, 
\begin{equation}
\langle \pi(q)|A_\mu^{\pi}|0 \rangle 
                         = -if_{\pi}q_\mu, \quad {\rm etc.}, 
                                                      \label{eq:PCAC}
\end{equation}
and 
\newpage
\begin{center}
\begin{quote}
{Table~I. Factorized amplitudes for two-body decays of charm mesons. 
The ellipses denote neglected contributions proportional to $f_-$.}
\end{quote}
\vspace{0.5cm}

\begin{tabular}
{l|l}
\hline\hline
$\quad\,\,${\rm Decay}
&\hskip 5cm {$\quad M_{\rm fact}\,$}
\\
\hline 
$D^+ \rightarrow \pi^+\bar K^0$
& 
\begin{tabular}{l}
{$\quad iU_{cs}U_{ud}{G_F \over\sqrt{2}}
a_1f_\pi(m_D^2-m_K^2)f_+^{\bar KD}(m_\pi^2)$}  \\
{\hskip 3cm $\times\Bigl\{1 + \Bigl({a_2\over a_1}\Bigr)
\Bigl({f_K\over f_\pi}\Bigr)
\Bigl({m_D^2-m_\pi^2\over m_D^2-m_K^2}\Bigr)
\Bigl({f_+^{(\pi D)}(m_K^2)\over f_+^{(\bar KD)}(m_\pi^2)}\Bigr)
\Bigr\}+\cdots$} 
\end{tabular}
\\
$D^0 \rightarrow \pi^+K^-$
& $\quad iU_{cs}U_{ud}{G_F \over\sqrt{2}}
a_1f_\pi(m_D^2-m_K^2)f_+^{\bar KD}(m_\pi^2)\,+\cdots$
\\
$D^0 \rightarrow \pi^0\bar K^0$
& $\quad iU_{cs}U_{ud}{G_F \over\sqrt{2}}\sqrt{1 \over 2}
a_2f_K(m_D^2-m_\pi^2)f_+^{\pi D}(m_K^2)\,+\cdots$
\\
$D_s^+ \rightarrow K^+\bar K^0$
& $\quad iU_{cs}U_{ud}{G_F \over\sqrt{2}}
a_2f_K(m_{D_s}^2-m_K^2)f_+^{KD_s}(m_K^2)\,+\cdots$
\\
\hline

$D^0 \rightarrow \pi^+\pi^-$
& $-iU_{cs}U_{us}
{G_F \over\sqrt{2}}a_1f_\pi(m_D^2 - m_\pi^2)f_+^{\pi D}(m_\pi^2) \, 
                                                            + \cdots $
\\
$D^0 \rightarrow \pi^0\pi^0$
&  \hskip 4cm $ 0 \qquad\qquad\qquad\quad  + \cdots $
\\
$D^+ \rightarrow \pi^+\pi^0$
& $\quad iU_{cs}U_{us}{G_F \over\sqrt{2}}
\sqrt{1\over 2}(a_1+a_2)f_\pi(m_d^2 - m_\pi^2)f_+^{\pi D}(m_\pi^2) 
+ \cdots $ 
\\
$D^0 \rightarrow K^0\bar K^0$
& \hskip 4cm 0
\\
$D^0 \rightarrow K^+K^-$
& $\quad iU_{cs}U_{us}{G_F \over {2}}
a_1f_{K}(m_D^2-m_K^2)f_+^{\bar KD}(m_K^2) + \cdots $
\\
$D^+ \rightarrow K^+\bar K^0$
& $\quad iU_{cb}U_{ud}{G_F \over\sqrt{2}}
a_1f_K(m_D^2-m_K^2)f_+^{\bar KD}(m_K^2) + \cdots $
\\
$D_s^+ \rightarrow \pi^+K^0$
& $-iU_{cs}U_{us}
{G_F \over\sqrt{2}}a_1f_\pi(m_{D_s}^2 - m_K^2)f_+^{KD_s}(m_\pi^2) \, 
                                                            + \cdots $
\\
$D_s^+ \rightarrow \pi^0K^+$
& $\quad iU_{cs}U_{us}
{G_F \over\sqrt{2}}
\sqrt{1 \over 2}a_2f_\pi(m_{D_s}^2 - m_K^2)f_+^{KD_s}(m_\pi^2) \, 
                                                            + \cdots $
\\
\hline\hline
\end{tabular}

\end{center}
\vspace{0.5cm}
\begin{eqnarray}
&&\langle \bar K(p')|V_\mu^{}|D(p)\rangle 
= (p+p')_\mu f_+^{\bar KD}(q^2) 
              + q_\mu f_-^{\bar KD}(q^2), \quad {\rm etc.}, 
                                                        \label{eq:KD}
\end{eqnarray}
as usual, where $q = p - p'$. Using these expressions of current 
matrix elements, we obtain the factorized amplitudes for two-body
decays of charm mesons listed in Table I, where terms proportional 
to $f_-$ are neglected since their coefficients are small or since 
$f_-$ is expected to be small for large values of its argument. 
It is seen that the factorized amplitudes for 
$D^0 \rightarrow \pi^0\pi^0$ and 
$D^0 \rightarrow \pi^0\bar K^0$ 
described by the color mismatched diagrams, 
$c \rightarrow (d\bar d)_1\,+\,u$ and 
$c \rightarrow (s\bar d)_1\,+\,u$, 
respectively, are much smaller (the color suppression) than those for 
the spectator decays (because of $|a_1| \gg |a_2|$). The factorized 
amplitude for $D_s^+ \rightarrow K^+\bar K^0$ is described by a sum 
of the color mismatched diagram and the so-called annihilation 
diagram in the weak boson mass $m_W \rightarrow \infty$ limit. 
The former is again proportional to $a_2$. The latter is proportional 
to $f_-^{KK}(m_{D_s}^2)$ and neglected. However the observed rates for 
these decays are not very small. The vanishing factorized amplitude 
for $D^0 \rightarrow K^0\bar K^0$ reflects a cancellation between two 
possible annihilation diagrams while the measured rate for this decay 
is a little smaller than the ordinary ones but not extremely
suppressed. To get rid of these problems, the factorization has been 
implemented by multiplying the amplitudes by phase factors arising
from final state interactions\cite{NRSX}. However, the final state 
interactions are realized by dynamical contributions of various 
hadron states. Therefore, we study explicitly the dynamical 
contributions of various hadrons as the non-factorizable amplitudes 
in the next section. 

\section{Non-factorizable amplitudes}

Now we study non-factorizable amplitudes for two-body decays of charm 
mesons, $P_1({p}) \rightarrow P_2({p'}) + P_3({q})$. As mentioned in
the previous section, we assume that they are dominated by dynamical 
contributions of various hadron states. Then they can be estimated by 
using a hard pseudo-scalar-meson approximation in the infinite
momentum frame (IMF, {\it i.e.}, ${\bf p} \rightarrow \infty$). 
It is an innovation of the old soft pion technique\cite{softpion}. 
The non-factorizable amplitude in this approximation is given 
by\cite{HARDP,suppl} 
\begin{equation}
M_{\rm non-f}(P_1\rightarrow P_2P_3)
\simeq M_{\rm ETC}(P_1\rightarrow P_2P_3)
+ M_{\rm S}(P_1\rightarrow P_2P_3). 
                                                      \label{eq:HP}
\end{equation} 
The equal-time commutator term ($M_{ETC}$) and the surface term
($M_S$) are given by 
\begin{equation}
M_{\rm ETC}(P_1\rightarrow P_2P_3)
= {i \over \sqrt{2}f_{P_3}}
     \langle{P_2|[V_{\bar P_3}, H_w]|P_1}\rangle 
                      + (P_2 \leftrightarrow P_3)   \label{eq:ETC}
\end{equation}
and
\begin{eqnarray} 
M_{\rm S}(P_1\rightarrow P_2P_3)
= {i \over \sqrt{2}f_{P_3}}
\Biggl\{&&\sum_n\Bigl({m_{P_2}^2 - m_{P_1}^2 
                                 \over m_n^2 - m_{P_1}^2}\Bigr)
  \langle{P_2|A_{\bar P_3}|n}\rangle
                         \langle{n|H_w|P_1}\rangle  \nonumber\\
&& + \sum_\ell\Bigl({m_{P_2}^2 - m_{P_1}^2 
                              \over m_\ell^2 - m_{P_2}^2}\Bigr)
\langle{P_2|H_w|\ell}\rangle
                  \langle{\ell|A_{\bar P_3}|P_1}\rangle\Biggr\} 
+ (P_2 \leftrightarrow P_3),  
                                                    \label{eq:SURF}
\end{eqnarray}
respectively, where $[V_\pi + A_\pi, H_w]=0$ has been used. (See 
Refs.\cite{HARDP} and \cite{suppl} for notations.) $M_{ETC}$ has the
same form as the one in the old soft pion approximation but now has 
to be evaluated in the IMF. The surface term has been given by a sum 
of all possible pole amplitudes, {\it i.e.}, $n$ and $\ell$ run over 
all possible single meson states, not only ordinary $\{q\bar q\}$, 
but also hybrid $\{q\bar qg\}$, four-quark $\{qq\bar q\bar q\}$ and 
glue-balls. Since the value of wave function of orbitally excited 
$\{q\bar q\}_{L \neq 0}$ state at the origin is expected to vanish in 
the non-relativistic quark model, or more generally, wave function 
overlappings between the ground-state $\{q\bar q\}_{L=0}$ and their 
excited states are expected to be small, however, we neglect 
contributions of these states. In the $u$-channel [the second line of 
the right-hand-side of Eq.(\ref{eq:SURF})], excited meson
contributions will be not very important because of 
$m_\ell^2 \gtrsim  m_{P_1}^2 \gg m_{P_2}^2$ if $\ell$ is an 
excited-state meson. In contrast, in the $s$ channel, we need to 
treat carefully contributions of exotic (non-$\{q\bar q\}$) mesons to 
charm decays (if they exist) since they have been predicted around 
charm masses. The $s$ channel of the color favored and mismatched 
spectator decays proceeds via four quark states after the weak 
interactions and therefore four-quark meson poles can contribute to 
these decays. However, in the annihilation decays in the weak boson 
mass $m_W \rightarrow \infty$ limit, their $s$ channel is given by 
$\{q\bar q\}$ state just after the weak interactions. Therefore we
expect that the ground-state $\{q\bar q\}_{0}$ and hybrid mesons can
give important $s$-channel pole contributions to these decays. 
(However, we neglect contributions of scalar hybrids in this article 
since their masses have been expected to be considerably lower than 
the charm ones\cite{Landua} and their contributions will be small.) 
The $s$-channel penguin can induce an $s$-channel pole contribution 
of glue-ball. In this way, the hard pseudo-scalar-meson amplitude in 
Eq.(\ref{eq:HP}) with Eqs.(\ref{eq:ETC}) and (\ref{eq:SURF}) as the 
non-factorizable contribution is described in terms of 
{\it asymptotic matrix elements} (matrix elements taken between 
single hadron states with infinite momentum) of charges $V_i$ and 
$A_i$, $(i=\pi\,\, {\rm and}\,\, K)$, and the effective weak 
Hamiltonian $H_w$. 

Asymptotic matrix elements of isospin and flavor $SU_f(3)$ charges, 
$V_{\pi}$ and $V_K$, are parameterized as 
\begin{eqnarray}
&&\langle{\pi^0|V_{\pi^+}|\pi^-}\rangle 
    = \sqrt{2}\langle{K^{+}|V_{\pi^+}|K^0}\rangle 
    = -\sqrt{2}\langle{D^{+}|V_{\pi^+}|D^0}\rangle \nonumber\\
&&\quad  = -2\langle{K^{+}|V_{K^+}|\pi^0}\rangle 
    = -\sqrt{2}\langle{D_s^{+}|V_{K^+}|D^0}\rangle 
    = \cdots = \sqrt{2}.                            \label{eq:MEV}
\end{eqnarray}
The above parameterization can be obtained by applying {\it asymptotic}
$SU_f(4)$ symmetry\cite{ASYMP} or $SU_f(4)$ extension of the nonet 
symmetry in $SU_f(3)$ to the matrix elements or by using the quark
counting. Matrix elements of axial counterpart, $A_\pi$ and $A_K$, of 
the above $V_\pi$ and $V_K$ are parameterized as 
\begin{eqnarray}
&&\langle{\rho^0|A_{\pi^+}|\pi^-}\rangle 
    = \sqrt{2}\langle{K^{*+}|A_{\pi^+}|K^0}\rangle 
    = -\sqrt{2}\langle{D^{*+}|A_{\pi^+}|D^0}\rangle \nonumber\\
&&\quad  = -2\langle{K^{*+}|A_{K^+}|\pi^0}\rangle 
    = -\sqrt{2}\langle{D_s^{*+}|A_{K^+}|D^0}\rangle 
    = \cdots = h                                     \label{eq:MEA}
\end{eqnarray}
in the same way as the above. The above parameterization reproduces 
well\cite{suppl,Takasugi} the observed decay rates for 
$D^* \rightarrow D\pi$ and $D^* \rightarrow D\gamma$. 

Next, we parameterize asymptotic matrix elements of $H_w$ using quark
counting\cite{TBD}. We expect that the factorization will not be a
good approximation to estimate asymptotic matrix elements of $H_w$ 
since one of the external states in these matrix elements contains 
only light quarks and weak interactions occur in a deep sea of soft
gluons where color degree of freedom of quarks will be compensated 
by soft gluons. Therefore, we forget the color degree of freedom of
quarks for the moment and count only their flavors (and hence 
connected quark-line diagrams). Now we review the procedure to 
parameterize the asymptotic matrix elements of $H_w$. To this, first, 
we rewrite the effective weak Hamiltonian in Eq.(\ref{eq:WH}) as 
\begin{equation}
H_w
= {G_F\over\sqrt{2}}
\Bigl\{c_-O_-^{(s'c)} + c_+O_+^{(s'c)} 
                + (penguin\,\,term) + h.c.\Bigr\},   \label{eq:WH+-}
\end{equation}
where 
\begin{equation}
O_\pm^{(s'c)} = O_1^{(s'c)}  \pm  O_2^{(s'c)} .     \label{eq:OPM}
\end{equation}
The four-quark operators $O_\pm^{(s'c)}$ belong to ${\bf 64}$ and 
${\bf 20^{\prime\prime}}$, respectively, of $SU_f(4)$ in its symmetry 
limit. The normal ordered four-quark operators $O_\pm^{(s'c)}$ can be 
expanded into a sum of products of (a) two creation operators in the 
left and two annihilation operators in the right, (b) three creation 
operators in the left and one annihilation operator in the right, (c) 
one creation operator in the left and three annihilation operators in 
the right, and (d) all (four) creation operators or annihilation
operators of quarks and anti-quarks. We associate (a)$-$(d) with
quark-line diagrams describing different types of matrix elements of
$O_\pm^{(s'c)}$. For (a), we utilize the two creation and 
annihilation operators to create and annihilate, respectively, 
the quarks and anti-quarks belonging to the meson states 
$|\{q\bar q\}\rangle$ and $\langle \{q\bar q\}|$ in the asymptotic 
matrix elements of $O_\pm^{(s'c)}$. For (b) and (c), we need to add 
a spectator quark or anti-quark to reach {\it physical} processes, 
$\langle {\{qq\bar q\bar q\}|O_\pm^{(s'c)}|\{q\bar q\}}\rangle$ and 
$\langle {\{q\bar q\}|O_\pm^{(s'c)}|\{qq\bar q\bar q\}}\rangle$,  
where $\{qq\bar q\bar q\}$ denotes four-quark mesons\cite{Jaffe}. 
They can be classified into the following four types, 
$\{qq\bar q\bar q\} = [qq][\bar q\bar q] \oplus (qq)(\bar q\bar q) 
\oplus \{[qq](\bar q\bar q) \pm (qq)[\bar q\bar q]\}$. 
Here () and [] denote symmetry and anti-symmetry, respectively, under
the exchange of flavors between them. Since only the first two can 
have $J^{P(C)}=0^{+(+)}$, we here consider contributions of 
them. These two types of four-quark mesons are again classified into
two different types with two different combinations of color degree 
of freedom, {\it i.e.}, one consists of color singlet $\{q\bar q\}$ 
pairs and the other consists of color octet $\{q\bar q\}$ pairs. They 
can mix with each other. Their mass eigen states are listed in the
tables in APPENDIX A. As seen in these tables, the predicted 
masses of the lighter components with relevant quantum numbers are 
much smaller than the charm mass while some of the heavier ones 
(with $\ast$ in the tables) can have masses close to the charm mass. 
We here take into account only contributions of the latter as an 
approximation. (For more precise arguments, we need all contributions 
of these exotic mesons.)

Counting all possible quark-line diagrams, we obtain sum rules which 
should be satisfied by asymptotic matrix elements of $O_\pm^{(s'c)}$. 
In this process, symmetry (or anti-symmetry) property of wave 
functions of external meson states under exchange of their quark and 
anti-quark plays an important role and therefore we have to be 
careful with the order of the quark(s) and anti-quarks(s) in 
$O_\pm^{(s'c)}$. Noting that the wave function of the ground-state 
$\{q\bar q\}_0$ meson is antisymmetric\cite{CLOSE} under the exchange 
of its quark and anti-quark, we obtain the following constraints on 
asymptotic matrix elements of $O_\pm^{(s'c)}$\cite{TBD}, 
\begin{eqnarray}
&&\langle{\{q\bar q\}_0|O_+^{(s'c)}|\{q\bar q\}_0}\rangle = 0, 
                                                \label{eq:SUM-G}\\
&&\langle{\,[qq]\,[\bar q\bar q]\,|O_+^{(s'c)}|\{q\bar q\}_0}\rangle
= \langle{\{q\bar q\}_0|O_+^{(s'c)}|[qq]\,[\bar q\bar q]\,}\rangle = 0, 
                                             \label{eq:SUM-anti}\\
&&\langle{(qq)(\bar q\bar q)|O_-^{(s'c)}|\{q\bar q\}_0}\rangle 
= \langle{\{q\bar q\}_0|O_+^{(s'c)}|(qq)(\bar q\bar q)}\rangle =0,  
                                                 \label{eq:SUM-sym}
\end{eqnarray}
and then, from them, we can obtain selection rules of asymptotic 
matrix elements of $H_w$. We summarize and parameterize them in 
APPENDIX B which have been given separately in Refs. \cite{TBD} and 
\cite{Unified}. Inserting the above parameterizations of asymptotic 
matrix elements of $A_\pi$, $A_K$ and $H_w$ into $M_S$ in 
Eq.(\ref{eq:SURF}), we obtain pole amplitudes including contributions 
of the $\{q\bar q\}_0$, $[qq][\bar q\bar q]$ and $(qq)(\bar q\bar q)$ 
mesons. 

A scalar glue-ball can give an important contribution, as an 
$s$-channel pole, to Cabibbo-angle suppressed decays through the 
$s$-channel penguin diagram. It can mix with scalar iso-singlet 
$\{q\bar q\}$ mesons. The glue-rich component of the mixture is 
described by $S^*$. We parameterize the ratio 
$\langle{K|A_K|S^*}\rangle$ to $\langle{\pi|A_\pi|S^*}\rangle$ by 
\begin{equation}
Z={\langle{K^+|A_{K^+}|S^*}\rangle 
\over \langle{\pi^+|A_\pi^+|S^*}\rangle},    \label{eq:Z}\\
\end{equation}
and then the residue of $S^*$ meson pole as 
\begin{equation}
\langle{K^+|A_{K^+}|S^*}\rangle 
\langle{S^*|H_w|D^0}\rangle
= -k_g\langle{\pi^+|H_w|D^+}\rangle .   \label{eq:para-g}
\end{equation}
Then the glue-ball contributions to the $D\rightarrow K\bar K$ and 
$D \rightarrow \pi\pi$ decays are given by 
\begin{eqnarray}
&&M_S^{(glue)}(D^0\rightarrow K^0\bar K^0) 
= {i \over f_K}\langle{\pi^+|H_w|D^+}\rangle
           \Biggl({m_D^2-m_K^2 \over m_D^2-m_{S^*}^2}\Biggr)2k_g \\
&&\hspace{3.9cm}\,= M_S^{(glue)}(D^0\rightarrow K^+K^-),           \\
&&M_S^{(glue)}(D^0\rightarrow \pi^0\,\,\pi^0\,)\,
= -{i \over f_\pi}\langle{\pi^+|H_w|D^+}\rangle
 {1\over Z} \Biggl({m_D^2-m_\pi^2 \over m_D^2-m_{S^*}^2}\Biggr)2k_g 
                                                                  \\
&&\hspace{3.9cm}\,
=\sqrt{{1\over 2}}M_S^{(glue)}(D^0\rightarrow \pi^+\pi^-).
\end{eqnarray}

An amplitude for a dynamical hadronic process can be decomposed into 
({\it continuum contribution}) + ({\it  Born term}).  
In the present case, $M_{\rm S}$ is given by a sum of pole amplitudes 
and therefore $M_{\rm ETC}$ corresponds to the continuum 
contribution \cite{MATHUR} which can develop a phase relative to the 
Born term. Therefore we here parameterize the ETC terms using isospin 
eigen amplitudes and their phases; {\it e.g.}, for the 
$M_{ETC}(D \rightarrow \pi\bar K)$'s, 
\begin{eqnarray}
&&M_{\rm ETC}(D^0\, \rightarrow \pi^+K^-) 
= \quad\,\,{1 \over 3}\,\,M_{\rm ETC}^{(3)}(D\rightarrow \pi\bar K)
                                        e^{i\delta_{3}(\pi\bar K)}\, 
+\,\, {2 \over 3}\,\,M_{\rm ETC}^{(1)}(D\rightarrow \pi\bar K)
                                        e^{i\delta_{1}(\pi\bar K)},  
                                                \\
&&M_{\rm ETC}(D^0\, \rightarrow \pi^0\,\bar K^0\,) \,
= -{\sqrt{2} \over 3}M_{\rm ETC}^{(3)}(D\rightarrow \pi\bar K)
                                        e^{i\delta_{3}(\pi\bar K)}\, 
+  {\sqrt{2} \over 3}M_{\rm ETC}^{(1)}(D\rightarrow \pi\bar K)
                                        e^{i\delta_{1}(\pi\bar K)},  
                                                \\
&&M_{\rm ETC}(D^+ \rightarrow  \pi^+\bar K^0)\, 
= \qquad\,\,\, M_{\rm ETC}^{(3)}(D\rightarrow \pi\bar K)
                                     e^{i\delta_{3}(\pi\bar K)}, 
\end{eqnarray}
where $M_{\rm ETC}^{(2I)}$'s are the isospin eigen amplitudes with
isospin $I$ and $\delta_{2I}$'s are the corresponding phase shifts
introduced\cite{phase}. $M_{ETC}$'s for decays into $K\bar K$ and 
$\pi\pi$ final states can be parameterized in a similar way. 
Since $M_{ETC}(D^0\rightarrow K^0\bar K^0)=0$, we obtain 
$M_{ETC}^{(0)}(D\rightarrow K\bar K)
=M_{ETC}^{(2)}(D\rightarrow K\bar K)$ and 
$\delta_0(K\bar K)=\delta_2(K\bar K)$, 
which lead to 
\begin{equation}
M_{ETC}(D^0 \rightarrow K^+K^-)=M_{ETC}(D^+ \rightarrow K^+\bar K^0)
= -{i\over \sqrt{2}f_K}
\langle{\pi^+|H_w|D^+}\rangle e^{i\delta_2(K\bar K)}. 
\end{equation}
The parameterization of $M_{ETC}(D \rightarrow \pi\pi)$ is taken to be 
compatible with that of the $D \rightarrow \pi\pi$ amplitudes in 
Ref. \cite{CLEO-pipi}. Because of the selection rule from 
Eq.(\ref{eq:SUM-G}), we obtain
\begin{eqnarray}
&&M_{ETC}^{(4)}(D\rightarrow\pi\pi)=0, \\
&&M_{ETC}^{(0)}(D\rightarrow\pi\pi)
={i\over \sqrt{2}f_\pi}\langle{\pi^+|H_w|D^+}\rangle
                                         e^{i\delta_0(\pi\pi)}, \\
&&M_{ETC}(D_s^+ \rightarrow K^+\bar K^0) 
=-{i\over \sqrt{2}f_K}\langle{\pi^+|H_w|D_s^+}\rangle
                                         e^{i\delta_2(K\bar K)} 
\end{eqnarray}
since the final state $K^+\bar K^0$ is of $I=1$. For the 
Cabibbo-angle favored $D\rightarrow \pi\bar K$ and suppressed 
$D_s^+\rightarrow (\pi K)^+$ decays, their ETC terms are given by 
\begin{eqnarray}
&& M_{ETC}^{(1)}(\,D\,\, \rightarrow \,\pi\,\bar K\,) 
= -{i\over \sqrt{2}f_\pi}\langle{\pi^+|H_w|D_s^+}\rangle
            \Bigl[2-{1\over 2}\Bigl({f_\pi\over f_K}\Bigr)\Bigr]
                              e^{i\delta_1(\pi\bar K)},          \\
&& M_{ETC}^{(3)}(\,D\,\, \rightarrow \,\pi\,\bar K\,) 
= -{i\over \sqrt{2}f_\pi}\langle{\pi^+|H_w|D_s^+}\rangle
                \Bigl[1-\Bigl({f_\pi\over f_K}\Bigr)\Bigr]
                              e^{i\delta_3(\pi\bar K)},          \\
&& M_{ETC}^{(1)}(D_s^+ \rightarrow (\pi K)^+) 
= -{i\over \sqrt{2}f_\pi}\langle{\pi^+|H_w|D^+}\rangle
               \Bigl[2-{1\over 2}\Bigl({f_\pi\over f_K}\Bigr)\Bigr]
                                  e^{i\delta_1(\pi K)},          \\
&& M_{ETC}^{(3)}(D_s^+ \rightarrow (\pi K)^+) 
= -{i\over \sqrt{2}f_\pi}\langle{\pi^+|H_w|D^+}\rangle
  \Bigl[1-\Bigl({f_\pi\over f_K}\Bigr)\Bigr]e^{i\delta_3(\pi K)}. 
\end{eqnarray}
In this way, the final state interactions are included in the 
non-factorizable amplitudes in the present perspective. It should be
noted that ETC terms for decays into exotic final states vanish (in 
the $SU_f(3)$ symmetry limit, {\it i.e.}, $f_K=f_\pi$). 

In the  approximation in which pole contributions of the ground-state 
$\{q\bar q\}_0$, scalar $[qq][\bar q\bar q]$ and $(qq)(\bar q\bar q)$ 
mesons and a glue-ball to $M_S$ are taken into account, we obtain the 
non-factorizable amplitudes for the 
$D \rightarrow \pi\bar K$, $\pi\pi$ and $K\bar K$ and 
$D_s^+ \rightarrow (K\bar K)^+$ and $(\pi K)^+$ decays 
inserting the above parameterizations of asymptotic matrix elements 
of charges and the effective weak Hamiltonian and the ETC terms 
implemented by the phase factors into Eq.(\ref{eq:HP}) with 
Eqs.(\ref{eq:ETC}) and (\ref{eq:SURF}). The result is listed 
in APPENDIX C. 

\section{Branching ratios}

Now we compare our result with experiments. To this, we need to know
values of parameters involved in our total amplitude (a sum of the 
factorized amplitude listed in Table~I and the corresponding hard 
pseudo-scalar-meson amplitude listed in APPENDIX C). We take
the following values of the CKM matrix elements and the decay 
constants\cite{PDG}; 
$U_{us}=-U_{cd}=0.21$, $U_{cs}=U_{ud}=0.98$ and $f_\pi=132$ MeV, 
$f_K=159$ MeV. 
To calculate decay branching ratios, we use the central values of 
the observed lifetimes of charm mesons\cite{PDG}; 
$\tau(D^+) = (1.057\pm 0.015)\times 10^{-12}$ s, 
$\tau(D^0) = (0.415\pm 0.004)\times 10^{-12}$ s and 
$\tau(D_s^+)=(0.467\pm 0.017)\times 10^{-12}$ s. 

The factorized amplitudes listed in Table~I contain the form factors, 
$f_+(q^2)$'s. We put $f_+(m_\pi^2)\simeq f_+(0)$ since $m_\pi^2$ is 
very small. The estimated values of $f_+(0)$'s are summarized in 
Ref.\cite{PDG} as 
\begin{eqnarray}
&&{\quad}f_+^{\bar KD}(0)_{expt}\,=\,0.75 \pm 0.02 \pm 0.02,  \\
&&\Biggl[{f_+^{\pi D}(0) \over f_+^{\bar KD}(0)}\Biggr]_{expt}
= 1.0 ^{+0.3}_{-0.2}\pm 0.4, \quad\qquad  [{\rm Mark\,\, III}],  \\
&&{\hspace{2.26cm}}\,\,
=\,1.3 \pm 0.2 \pm 0.1,  \quad\qquad\, [{\rm CLEO}].
\end{eqnarray}
The above result is compatible with the $SU_f(3)$ symmetry and 
therefore we assume that 
\begin{equation}
f_+^{\bar KD}(m_\pi^2)\simeq f_+^{\pi D}(m_\pi^2)
\simeq f_+^{KD_s}(m_\pi^2)\simeq f_+^{\pi D_s}(m_\pi^2)
\simeq f_+^{\bar KD}(0).  
                                                  \label{eq:FF}
\end{equation}
In this way, we can obtain the factorized amplitudes which provide 
the branching ratios $B_{fact}$'s in Table~II. It is seen that the 
branching ratios for the so-called spectator decays, 
$D^+\rightarrow \pi^+\pi^0$ and 
$D^+\rightarrow \pi^+\bar K^0$, 
are too big and that the color suppressed decays, 
$D^0\rightarrow \pi^0\pi^0$ and 
$D^+\rightarrow \pi^0\bar K^0$, 
are too strongly suppressed as mentioned before. 

Now we evaluate the non-factorizable amplitudes listed in APPENDIX C. 
The asymptotic matrix elements of  $A_\pi$ and $A_K$ which have been 
parameterized in Eq.(\ref{eq:MEA}) are estimated to be 
$|h|\simeq 1.0$\cite{HARDP,suppl} by using partially conserved 
axial-vector current (PCAC) and the observed rate\cite{PDG}, 
$\Gamma(\rho \rightarrow \pi\pi)_{expt} \simeq 150$ MeV. 

We here assign $f_J(1710)$ to the glue-rich scalar $S^*$ and take 
$\Gamma_{f_J}=175 \pm 9$ MeV\cite{PDG} as the width $\Gamma_{S^*}$ 
of $S^*$. Then $|\langle{K|A_K|S^*}\rangle|$ can be estimated from 
the observed width of $f_J$ and the ratio of its partial widths 
\begin{equation}
\Biggl[{\Gamma(f_J\rightarrow \pi\pi)
\over \Gamma(f_J\rightarrow K\bar K)}\Biggr]_{expt}=0.39 \pm 0.14
                                                \label{eq:RATIO}
\end{equation}
as 
$|\langle{K^+|A_{K^+}|S^*}\rangle| \lesssim 0.15$ 
by using PCAC. They are really small as expected. It implies that
overlapping between wave functions of the ground-state-meson and the
glue-ball (or glue-rich meson) is very small. Therefore size of 
matrix element of $H_w$ taken between $S^*$ and $K$ will be much 
smaller than that between $\pi$ and $K$, and hence the size of $k_g$ 
will be much smaller than unity. $Z$ in Eq.(\ref{eq:Z}) also can be 
estimated to be $Z \sim 2.0$\cite{f_0} from the above value of the 
observed ratio of decay rates, Eq.(\ref{eq:RATIO}), where we have
taken the same sign between $\langle{K^+|A_{K^+}|S^*}\rangle $ and 
$\langle{\pi^+|A_\pi^+|S^*}\rangle$ as expected in the $SU_f(3)$ 
symmetry. 

Although existence of four-quark mesons has never been confirmed, 
indications of their existence are increasing\cite{exotics}. We here 
take the predicted values of $[qq][\bar q\bar q]$ meson masses listed
in Table A1(a) in APPENDIX A. However, the predicted 
$(qq)(\bar q\bar q)$ meson masses listed in Table A2(a) satisfy 
$m_{E_{\pi\pi}^*} < m_D < m_{E_{\pi K}^*}$. 
In this case, it is hard that the $(qq)(\bar q\bar q)$ meson 
pole amplitudes for the spectator decays, 
$D^+\rightarrow \pi^+\bar K^0$ and $D^+\rightarrow \pi^+\pi^0$, 
interfere destructively with their factorized amplitudes, 
simultaneously. [The naively factorized amplitudes have lead to too
big rates for the spectator decays as discussed before.] Since the 
predicted values of their masses still would have ambiguities,
however, we here shift up the mass values of $(qq)(\bar q\bar q)$ 
mesons by 100 MeV from the predicted ones to obtain 
$m_D < m_{E_{\pi\pi}^*} < m_{E_{\pi K}^*}$ 
which leads to a destructive interference between the factorized
amplitudes and the four-quark meson pole amplitudes for the 
$D^+\rightarrow \pi^+\bar K^0$ and $D^+\rightarrow \pi^+\pi^0$.  
Masses of four-quark mesons containing a charm quark listed in Tables
A1(b) and A2(b) are estimated {\it crudely} by using the quark 
counting since our result is not very sensitive to them. Although 
widths of four-quark mesons are also still not known, the 
$[qq][\bar q\bar q]$ mesons with $*$ are expected to be narrower than 
the corresponding $(qq)(\bar q\bar q)$ mesons\cite{Jaffe}. 

For the phases $\delta_{2I}$'s arising from contributions of 
non-resonant multi-hadron intermediate states with isospin $I$,  
they will be restricted in the region $|\delta_{2I}| < 90^\circ$. 
(Resonant contributions have already been extracted as pole 
amplitudes in $M_S$.) We here treat them as adjustable parameters 
which are restricted in the region $|\delta_{2I}| < 90^\circ$ as 
mentioned above and satisfy the relations 
\begin{equation}
\delta_0(\pi\pi)=\delta_0(K\bar K)=\delta_2(K\bar K)
=\delta_1(\pi\bar K)=\delta_1(\pi K)\equiv \delta. 
                                                   \label{eq:phases}
\end{equation}
The second equality has been obtained previously and the others are 
expected in the $SU_f(3)$ symmetry limit. However the exotic phase 
shift $\delta_3(\pi\bar K)$ is treated as a parameter independent of 
non-exotic $\delta_1(\pi\bar K)$.  

The remaining parameters are $k_a^*$, $k_s^*$ and $k_g$ which provide 
the residues of poles of $[qq][\bar q\bar q]$, $(qq)(\bar q\bar q)$ 
and glue-ball, respectively. Since overlappings between wave 
functions of these exotic mesons and the ground-state ones are 
expected to be very small, values of these parameters will be much
smaller than unity, {\it i.e.}, $|k_g|,\,|k_a^*|,\,|k_s^*|\ll 1$.
It was predicted\cite{Jaffe} that the couplings of the four-quark 
$[qq][\bar q\bar q]$ and $(qq)(\bar q\bar q)$ mesons with $\ast$ to 
the ground-state $O^{-(+)}$ mesons would be small because of the 
structure of their wave functions with respect to spin and color 
degree of freedom and that the $(qq)(\bar q\bar q)$ mesons with 
$\ast$ would couple to the ground-state $O^{-(+)}$ mesons much more 
weakly than the $[qq][\bar q\bar q]$ mesons. The above statement 
implies that $|k_s^*| \ll |k_a^*|$. 

For values of the asymptotic ground-state-meson matrix elements of 
$H_w$, $\langle{\pi^+|H_w|D_s^+}\rangle$, etc., we have no
information. Therefore we here treat the above matrix element in
addition to the parameters mentioned above as adjustable parameters 
and look for overall fits to the observed branching ratios for two 
body decays, where the latter parameters are not perfectly free but 
restricted as discussed above, {\it i.e.}, 
$|k_g|\ll 1$, $|k_s^*| \ll |k_a^*| \ll 1$ 
and 
$|\delta|$, $|\delta_{exotic}| < 90^\circ$, 
where $\delta$ is given in Eq.(\ref{eq:phases}) and $\delta_{exotic}$ 
is the strong phase in the exotic $\pi K$ (or $\bar K$) channel, 
$\delta_{exotic}=\delta_{3}(\pi K)=\delta_{3}(\pi\bar K)$. 

We can reproduce remarkably well the observed branching 
ratios for Cabibbo-angle favored and suppressed two body decays of 
charm mesons, simultaneously, by taking reasonable values of
parameters involved, {\it i.e.}, very small values of parameters 
providing the residues of poles of glue-ball, $[qq][\bar q\bar q]$ 
and $(qq)(\bar q\bar q)$ mesons, 
$|k_g| \sim 0.1$, $|k_a^*| \sim 0.1$ and $|k_s^*| \sim 0.01$,
respectively, the predicted mass values of $[qq][\bar q\bar q]$ 
mesons in Ref.\cite{Jaffe} but $(qq)(\bar q\bar q)$ meson masses
larger by $100$ MeV than the predicted ones, and their widths, 
relatively narrower $\Gamma_{[qq][\bar q\bar q]}\sim 0.2$ GeV and 
rather broader $\Gamma_{(qq)(\bar q\bar q)}\sim 0.4$ GeV. 
These are compatible with the discussions in Ref.\cite{Jaffe}. 
For the phases of the ETC term relative to the surface term in 
the narrow width limit, which are expected to arise from 
contributions of multi hadron intermediate states, rather large 
values ($\gtrsim 70^\circ$) of these phases in Eq.(\ref{eq:phases}) 
are favored while our result is not very sensitive to the value of 
the exotic phases $\delta_3(\pi K)$ and $\delta_3(\pi\bar K)$ since 
contributions of the ETC terms into exotic final states are small 
(vanishing in the $SU_f(3)$ symmetry limit, {\it i.e.}, $f_K=f_\pi$). 
For the asymptotic ground-state-meson matrix element of $H_w$, 
Cabibbo-angle favored and suppressed ones separately satisfy the charm 
counterparts of the (asymptotic) $\Delta I={1\over 2}$ rule in the
strangeness changing hadronic weak interactions of $K$ mesons, and are 
related to each other by using (asymptotic) $SU_f(3)$ symmetry as
discussed in APPENDIX B. Therefore we here consider only the matrix 
element of the Cabibbo-angle favored Hamiltonian,  
$\langle{\pi^+|H_w|D_s^+}\rangle$, which favors values not very 
far from 
\begin{eqnarray}
&&\langle{\pi^+|H_w|D_s^+}\rangle_{\rm fact} 
={G_F\over \sqrt{2}}U_{ud}U_{cs}
\Bigl\{{m_\pi^2+m_{D_s}^2}\Bigr\}f_\pi f_{D_s}c_-
\simeq  0.078\times 10^{-5}\,\,{\rm (GeV)}^2  
                                                   \label{eq:AMHw}
\end{eqnarray}
obtained by using a factorization compatible with the quark counting 
discussed before. In the above estimate of 
$\langle{\pi^+|H_w|D_s^+}\rangle_{\rm fact}$, we have taken 
$c_-=0.885$ (and $c_+=0.375$) with the leading order QCD 
corrections\cite{NRSX} and a recent lattice result on the decay 
constant, $f_{D_s} \simeq 216$ MeV\cite{Okawa}. It may imply that 
our estimate of asymptotic matrix element of $H_w$ is reasonable (but 
it does not necessarily imply that the factorization is a reliable 
approximation). 

As an example, a typical result is shown in Table~II where we have
taken the following values of unknown parameters; the asymptotic 
matrix element of $H_w$, 
$\langle{\pi^+|H_w|D_s^+}\rangle 
\simeq 0.0695\times 10^{-5}$ (GeV)$^2$, 
which is not very far from the above 
$\langle{\pi^+|H_w|D_s^+}\rangle_{\rm fact}$, 
the strong phases, 
{$\delta_0(\pi\pi)=\delta_1(\pi\bar K)=\delta_1(\pi K)
=\delta_3(\pi\bar K)=\delta_3(\pi K)=85^\circ$},  
and the residues of meson poles, 
$k_0=0.71$, $k_a^*=-0.20$, $k_s^*=-0.02$, $k_g=0.085$. 
It is seen, from Table~II, that the naively factorized amplitudes for 
the so-called spectator decays which lead to too big rates for these 
decays interfere destructively with the exotic meson pole amplitudes 
as expected. The non-factorizable amplitudes can supply significant 
contributions to the so-called color suppressed 
$D^0 \rightarrow \pi^0\bar K^0$ and $D^0 \rightarrow \pi^0\pi^0$
which are strongly suppressed in the naive factorization. To solve 
the well-known puzzle that the observed ratio of rates for 
$D^0\rightarrow K^+K^-$ to $D^0\rightarrow \pi^+\pi^-$ is around 
2.5, a peculiar $SU_f(3)$ symmetry breaking may have to be 
introduced\cite{Chau}. In the present case, such a symmetry breaking 
can be realized dominantly by different contributions of 
$[qq][\bar q\bar q]$ meson poles, {\it i.e.}, 
$m_D^2-m_{\hat \sigma^*}^2 \gg m_D^2-m_{\hat \sigma^{s*}}^2$. 
It is known that the $D^0 \rightarrow K^0\bar K^0$ decay is described 
by two annihilation diagrams (in the $m_W \rightarrow \infty$ limit) 
which cancel each other. Therefore both the factorized and the 
non-factorizable amplitudes for 
\newpage
\begin{quote}
{Table~II. Branching ratios ($\%$) for two-body decays of charm
mesons where $a_1=1.09$ and $a_2=-0.09$ have been used. Values of 
parameters introduced are tentatively taken as follows; 
the matrix element 
$\langle{\pi^+|H_w|D_s^+}\rangle 
\simeq 0.0695\times 10^{-5}$ (GeV)$^2$, 
the phases 
{$\delta_0(\pi\pi)=\delta_1(\pi\bar K)=\delta_1(\pi K)
=\delta_3(\pi\bar K)=\delta_3(\pi K)=85^\circ$}, 
and the parameters providing the residues of various meson poles 
{$k_0=0.71$, $k_a^*=-0.20$, $k_s^*=-0.02$, $k_g=0.085$}. 
$B_{\rm fact}$, $B_{\rm non-f}$ and $B_{\rm total}$ include only the
factorized amplitude, only the non-factorizable one (involving the 
$\{q\bar q\}_0$, $[qq][\bar q\bar q]$ and $(qq)(\bar q\bar q)$ meson
poles) and the sum of them, respectively.  
The data values are taken from Ref.\cite{BHP}.} 
\end{quote}

\begin{center}
\begin{tabular}
{|l|c|c|c|c|}
\hline
{$\quad\,\,${\rm Decay}}
&$\quad B_{\rm fact}\quad$
&{$B_{\rm non-f}$} 
&$\quad B_{total}\quad$
&$\quad B_{\rm expt}\quad$\\
\hline
$\quad D^+ \rightarrow \pi^+\bar K^0\quad$
&{ 9.66 }
&
{\quad 2.15\,\quad} 
& {  $2.71$ }
& {  $2.74 \pm 0.29$}
\\
\hline
$\quad D^0\, \rightarrow \pi^+K^-\quad$
&{  4.66 }
&{ 6.75 } 
& { 3.91 } 
& {  $3.83 \pm 0.12$ }
\\
\hline
$\quad D^0\, \rightarrow \pi^0\bar K^0\quad$
&{ 0.03 } 
&{ 1.94}
& { 2.26 } 
& {$2.11 \pm 0.21$}
\\
\hline
$\quad D_s^+ \rightarrow K^+\bar K^0\quad$
& { 0.06 }
& { 3.82 }
& { 2.93 } 
& { $3.6 \pm 1.1$ }
\\
\hline
$\quad D^0\, \rightarrow \pi^+\pi^-\quad$
& { 0.29 }
& { 0.68 }
& { 0.14 } 
& { $0.152 \pm 0.011$ }
\\
\hline
$\quad D^0\, \rightarrow \pi^0\pi^0\quad$
& { 0.00 }
& { 0.05 }
& { 0.05 }
& { $0.084 \pm 0.022$ }
\\
\hline
$\quad D^+ \rightarrow \pi^+\pi^0\quad$
& { 0.31 }
& {  1.11 }
& { 0.25 }
& { $0.25 \pm 0.07$ }
\\
\hline
$\quad D^0\, \rightarrow K^+K^-\quad$
& { 0.32 }
& { 0.19 }
& { 0.41 }
& { $0.43 \pm 0.03$ }
\\
\hline
$\quad D^0\, \rightarrow K^0\bar K^0\quad$
& { 0.00 }
& { 0.12 }
& { 0.12 }
& { $0.13 \pm 0.04$ }
\\
\hline
$\quad D^+ \rightarrow K^+\bar K^0\quad$
& { 0.81 }
& { 1.29 }
& { 0.98 }
& { $0.72 \pm 0.12$ }
\\
\hline
$\quad D_s^+ \rightarrow \pi^+K^0\quad$
& { 0.32 }
& { 0.14 }
& { 0.22 }
& { $<0.70$ }
\\
\hline
$\quad D_s^+ \rightarrow \pi^0 K^+\quad$
& { 0.00 }
& { 0.02 }
& { 0.02 }
& { $<0.70$ }
\\
\hline
\end{tabular}

\end{center}
\vspace{0.5 cm}
this decay vanish. However the $s$-channel penguin can induce a pole 
contribution of scalar glue-ball or glue-rich scalar meson which 
leads to a reasonable size of rate for this decay in consistency with 
the $D^0\rightarrow \pi\pi$ and $K^+K^-$ decays.  

\section{Summary}

Two body decays of charm mesons have been studied by describing their
amplitude in terms of a sum of factorized and non-factorizable ones. 
The former has been estimated by using the naive factorization in the
BSW scheme while the latter has been calculated by using a hard 
pseudo-scalar-meson approximation. It has been given by a sum of the
equal-time commutator term and the surface term which contains pole 
contributions of various meson states, not only the ground-state 
$\{q\bar q\}_0$ mesons but also a glue-ball and exotic 
$\{qq\bar q\bar q\}$ mesons with $J^{P(C)}=0^{+(+)}$. 

In this way, a possible solution to the long standing problems in 
charm meson decays has been given, at least, qualitatively. 
Factorizable contributions which lead to too big branching ratios 
$B_{\rm fact}$'s for the spectator decays like 
$D^+ \rightarrow \pi^+\pi^0$ and $\pi^+\bar K^0$ can interfere
destructively with non-factorizable ones and a sum of these two
contributions can reproduce their observed values of branching
ratios. The naive factorization also leads to too strong color 
suppression. However, non-factorizable amplitudes can supply 
sufficient contributions to the color suppressed decays. 
The observed branching ratios for the mixed decays which have both 
contributions from the spectator and the color mismatched diagrams 
can be reproduced by interferences between the factorizable and the 
non-factorizable contributions. 

Two body decays of charm mesons into final states including $\eta$ or 
$\eta'$, in particular, decays into $\pi\eta$ and $\pi\eta'$ are 
interesting. However non-factorizable contributions to these decays 
are complicated\cite{pi-eta} because of the $\eta$-$\eta'$ mixing and 
therefore these decays should be investigated separately. 

For quasi two body $D \rightarrow VP$ decays, the mixing between 
iso-singlet mesons in the final states is rather simple, {\it i.e.}, 
the $\omega$-$\phi$ mixing is known to be approximately 
ideal. However, in these decays,  all the four types of four-quark 
$\{qq\bar q\bar q\} = [qq][\bar q\bar q] \oplus (qq)(\bar q\bar q) 
\oplus \{[qq](\bar q\bar q) \pm (qq)[\bar q\bar q]\}$ 
mesons can contribute except for annihilation decays. Fortunately, 
the decays, 
$D^0 \rightarrow \bar K^0\phi$, $D_s^+ \rightarrow \pi^+\rho^0$ and 
$\pi^+\omega$, 
are described approximately by annihilation diagrams and four-quark 
meson contributions can be neglected. Here the first one has been
observed with a substantial rate but the last two are suppressed. 
Since the factorized amplitudes for these decays are always 
suppressed\cite{BSW}, the observed rates should be dominantly 
supplied by non-factorizable dynamical contributions of various
hadrons in the present approach. The $D^0 \rightarrow \bar K^0\phi$ 
amplitude is dominantly given by a sum of its ETC term describing 
contributions of multi hadron intermediate states and the $\bar K^0$ 
meson pole amplitude\cite{Kbar-phi}. The observed suppression of the 
$D_s^+ \rightarrow \pi^+\rho^0$ suggests that a hybrid pseudo scalar 
meson ($\pi_H$) with a mass very close to $m_{D_s}$ and with a rather 
narrow width exists (a recently observed pseudo scalar hybrid meson 
$\pi(1800)$\cite{Prokoshkin} may be assigned to this one although 
its mass $\sim 1.8$ GeV is not sufficiently close to $m_{D_s}$) and 
that its pole contribution cancel a sum of the ETC term and the pion 
pole amplitude for this decay\cite{pi-rho}. A strange component 
($K_H$) belonging to the same multiplet as $\pi_H$ does not disturb
our good result on the $D^0 \rightarrow \bar K^0\phi$ decay. 
A recent observation of the $D_s^+ \rightarrow \pi^+\omega$ with a 
small rate\cite{pi-omega}, 
$B(D_s^+\rightarrow\omega\pi^+)=(2.7 \pm 1.2)\times 10^{-3}$,  
suggests that an iso-triplet hybrid meson with $J^{PC}=1^{+-}$ exists 
but is not very close to $m_{D_s}$ (probably much lower than 
$m_{D_s}$ as expected\cite{Landua}) or couples very weakly to 
$\pi\omega$. Quasi two body decays of charm mesons including 
factorizable contributions will be investigated more extensively 
elsewhere. 

Finally, hadronic weak interactions of charm mesons are intimately 
related to hadron spectroscopy. More informations of hadron
spectroscopy will be needed to find a more quantitative solution to
the puzzles in hadronic weak interactions. 
 
\vspace{1cm}

\thanks{
The author would like to thank to Prof. T.~E.~Browder and the other 
members of high energy physics group of University of Hawaii for 
their discussions, comments and hospitality during his stay there. 
}

\newpage
\appendix
\section{Four-quark mesons}
Scalar four-quark mesons with charm quantum number $C=0$ and 1 are 
listed, where $S$ and $I$ denote the strangeness and isospin. 
Particles with superscript $^s$ contain an $s\bar s$ pair. Mass 
values of non-charm ($C=0$) mesons are given in Ref.\cite{Jaffe}. 
Masses of four-quark charm ($C=1$) mesons estimated by using the 
quark counting (with $m_u=m_d=0$, $m_s=0.2$ GeV, $m_c=1.5$ GeV and 
the predicted mass values of $C=0$ mesons mentioned above) are given 
between parentheses ( ). Particles containing double $(s\bar s)$ 
pairs and $(c\bar c)$ pair(s) are dropped since they do not 
contribute in this paper. 

\begin{center}
\begin{quote}
{Table A1(a). Ideally mixed scalar $[qq][\bar q\bar q]$ mesons with
$C=0$.} 
\end{quote}
\vspace{0.5cm}

\begin{tabular}
{|c|c|c|c|c|}
\hline
$\quad S\quad$
&$\quad I=1\quad$
&$\quad I={1\over 2}\quad$
&$\quad I=0\quad$
&$\quad$Mass(GeV)$\quad$
\\
\hline
$1$
&
&
\begin{tabular}{c}
{$\hat\kappa$}\\
{$\hat\kappa^*$}\\
\end{tabular}
&
&
\begin{tabular}{c}
{$0.90$}\\
{$1.60$}\\
\end{tabular}
\\
\hline
0
&
\begin{tabular}{c}
{}\\
{}\\
{$\hat\delta^s$}\\
{$\hat\delta^{s*}$}\\
\end{tabular}
&
&\begin{tabular}{c}
{$\hat\sigma$}\\
{$\hat\sigma^*$}\\
{$\hat\sigma^s$}\\
{$\hat\sigma^{s*}$}\\
\end{tabular}
&\begin{tabular}{c}
{$0.65$}\\
{$1.45$}\\
{${1.10}$}\\
{${1.80}$}\\
\end{tabular}

\\
\hline
\end{tabular}

\end{center}
\begin{center}
\begin{quote}
{Table A1(b). Ideally mixed scalar $[qq][\bar q\bar q]$ mesons with
$C=1$.} 
\end{quote}
\vspace{0.5cm}

\begin{tabular}
{|c|c|c|c|c|}
\hline
$\quad S\quad$
&$\quad I=1\quad$
&$\quad I={1\over 2}\quad$
&$\quad I=0\quad$
&$\quad$Mass(GeV)$\quad$
\\
\hline
$1$
&
\begin{tabular}{c}
$\hat F_I$ \\
$\hat F_I^*$
\end{tabular}
&
&
\begin{tabular}{c}
$\hat F_0$\\
$\hat F_0^*$
\end{tabular}
&
\begin{tabular}{c}
{(2.4)}\\
{(3.2)}
\end{tabular}
\\
\hline
0
&
&
\begin{tabular}{c}
{$\hat D$}\\
{$\hat D^*$}\\
{$\hat D^s$}\\
{$\hat D^{s*}$}
\end{tabular}
&
&\begin{tabular}{c}
{(2.2)}\\
{(3.0)}\\
{(2.6)}\\
{(3.4)}
\end{tabular}

\\
\hline 
\end{tabular}

\end{center}
\begin{center}
\begin{quote}
{Table A2(a). Ideally mixed scalar $(qq)(\bar q\bar q)$ mesons with
$C=0$.}

\end{quote}

\vspace{0.5cm}

\begin{tabular}
{|c|c|c|c|c|c|c|}
\hline
$\quad S\quad$
&$\quad I=2\quad$
&$\quad I={3\over 2}\quad$
&$\quad I=1\quad$
&$\quad I={1\over 2}\quad$
&$\quad I=0\quad$
&$\,\,$ Mass(GeV) $\,\,$
\\
\hline
$2$
&
&
&
\begin{tabular}{c}
{$E_{KK}$}\\
{$E_{KK}^*$}
\end{tabular}
&
&
&
\begin{tabular}{c}
{$1.55$}\\
{$2.10$}
\end{tabular}
\\
\hline
1
&
&
\begin{tabular}{c}
{$E_{\pi K}$}\\
{$E_{\pi K}^*$} \\
{}\\
{}\\
\end{tabular}
&
&
\begin{tabular}{c}
{$C_K$}\\
{$C_K^*$}\\
{$C_K^s$}\\
{$C_K^{s*}$}\\
\end{tabular}
&
&
\begin{tabular}{c}
{$1.35$}\\
{$1.95$}\\
{$1.75$}\\
{$2.20$}\\
\end{tabular}
\\
\hline
0
&
\begin{tabular}{c}
{$E_{\pi\pi}$}\\
{$E_{\pi\pi}^*$} \\
{}\\
{}\\
\end{tabular}
&
&
\begin{tabular}{c}
{$C_\pi$}\\
{$C_\pi^*$} \\
{$C_\pi^s$}\\
{$C_\pi^{s*}$}\\
\end{tabular}
&
&
\begin{tabular}{c}
{$C$}\\
{$C^*$}\\
{$C^s$}\\
{$C^{s*}$}\\
\end{tabular}
&
\begin{tabular}{c}
{$1.15$}\\
{$1.80$}\\
{$1.55$}\\
{$2.10$}\\
\end{tabular}
\\
\hline
\end{tabular}

\end{center}
\vspace{0.5cm}
\begin{center}
\begin{quote}
{Table A2(b). Ideally mixed scalar $(qq)(\bar q\bar q)$ mesons with
$C=1$.}

\end{quote}

\vspace{0.5cm}

\begin{tabular}
{|c|c|c|c|c|c|c|}
\hline

$\quad S\quad$
&$\quad I=2\quad$
&$\quad I={3\over 2}\quad$
&$\quad I=1\quad$
&$\quad I={1\over 2}\quad$
&$\quad I=0\quad$
&$\,\,$ Mass(GeV) $\,\,$
\\
\hline
$2$
&
&
&
&
\begin{tabular}{c}
{$E_{KF}$}\\
{$E_{KF}^*$}
\end{tabular}
&
&
\begin{tabular}{c}
{(3.1)}\\
{(3.7)}
\end{tabular}
\\
\hline
1
&
&
&
\begin{tabular}{c}
{$E_{\pi F}$}\\
{$E_{\pi F}^*$}\\
{}\\
{}\\
\end{tabular}
&
&
\begin{tabular}{c}
{$C_F$}\\
{$C_F^*$}\\
{$C_F^s$}\\
{$C_F^{s*}$}\\
\end{tabular}
&
\begin{tabular}{c}
{(2.9)}\\
{(3.5)}\\
{(3.3)}\\
{(3.9)}\\
\end{tabular}
\\
\hline
0
&
&
\begin{tabular}{c}
{$E_{\pi D}$}\\
{$E_{\pi D}^*$}\\
{}\\
{}\\
\end{tabular}
&
&
\begin{tabular}{c}
{$C_D$}\\
{$C_D^*$} \\
{$C_D^s$}\\
{$C_D^{s*}$}\\
\end{tabular}
&
&
\begin{tabular}{c}
{(2.7)}\\
{(3.3)}\\
{(3.1)}\\
{(3.7)}\\
\end{tabular}
\\
\hline
$-1$
&
&
&
\begin{tabular}{c}
{$E_{\bar KD}$}\\
{$E_{\bar KD}^*$}
\end{tabular}
&
&
&
\begin{tabular}{c}
{(2.9)}\\
{(3.5)}
\end{tabular}
\\
\hline
\end{tabular}

\end{center}
\vspace{1cm}
\section{Asymptotic matrix elements of $H_w$}

Constraints on asymptotic matrix elements of $H_w$ have been
previously derived by using an algebraic method based on commutation 
relations between charges and currents (and hence the effective weak
Hamiltonian) and, then, by counting all possible connected quark-line 
diagrams. We here summarize a part of them which are useful in
this paper. We here describe the Cabibbo-angle favored 
($\Delta C=-1,\,\Delta S=-1$) and suppressed 
($\Delta C=-1,\,\Delta S=0$) weak Hamiltonians as $H_w(-,-)$ and 
$H_w(-,0)$, respectively, for convenience' sake. 

\hfil\break
(i) Constraints on asymptotic ground-state-meson matrix elements of
$H_w$:
\begin{equation}
\left\{ 
\begin{array}{l}
\langle{\bar K^0\,\,|H_w(-,-)|D^0}\rangle 
= - \langle{\pi^+|H_w(-,-)|D_s^+ }\rangle,      \\
\langle{\bar K^{*0}|H_w(-,-)|D^0}\rangle 
=-\langle{\rho^+|H_w(-,-)|D_s^+ }\rangle           \\
= \langle{\bar K^{0}|H_w(-,-)|D^{*0}}\rangle 
=-\langle{\pi^+|H_w(-,-)|D^{*0}}\rangle                   
= (\sqrt{2}k_0/ h)\langle{\bar K^0|H_w(-,-)|D^0}\rangle,
\end{array} 
\right.
\end{equation}
\begin{equation}
\left\{ 
\begin{array}{l}
\langle{\pi^+|H_w(-,0)|D^+}\rangle \,\,
= - \langle{K^+|H_w(-,0)|D_s^+ }\rangle,      \\
\langle{\pi^{+}|H_w(-,0)|D^{*+}}\rangle 
=- \langle{K^{+}|H_w(-,0)|D_s^{*+}}\rangle     \\       
=\langle{\rho^+|H_w(-,0)|D^+ }\rangle      
=-\langle{K^{*+}|H_w(-,0)|D_s^{+}}\rangle 
= (\sqrt{2}k_0/h)\langle{\pi^+|H_w(-,0)|D^+}\rangle,
\end{array} 
\right.
\end{equation}
where $k_0=\sqrt{1\over 2}h$ with 
$h=\langle{\rho^0|A_{\pi^+}|\pi^-}\rangle$ 
has been obtained by using an algebraic method\cite{Tanuma}. 
It will be understood more intuitively since all the external states 
in the above matrix elements of $H_w$ are of helicity = 0 states of 
the ground-state $\{q\bar q\}_0$ mesons and the difference of spins 
will be not very important in the IMF. 
\newpage
\hfil\break
(ii) Constraints on asymptotic matrix elements of $H_w$ between the
ground-state-meson and $[qq][\bar q\bar q]$ meson states:
\begin{equation}
\left\{ 
\begin{array}{l}
\langle{\hat{\bar\kappa}^{*0}|H_w(-,-)|D^0}\rangle 
=-\langle{\delta^{s*+}|H_w(-,-)|D_s^+ }\rangle 
= (k_a^*/2 A_a^*)\langle{\bar K^0|H_w(-,-)|D^0}\rangle, \\
\langle{\bar K^0|H_w(-,-)|\hat D^{*0}}\rangle 
=-\langle{\bar K^0|H_w(-,-)|\hat D^{s*+}}\rangle \\
=\sqrt{1\over 2}\langle{\pi^+|H_w(-,-)|\hat F_I^{*+}}\rangle 
=-\sqrt{1\over 2}\langle{\pi^0|H_w(-,-)|\hat F_I^{*0}}\rangle \\
\hspace{8cm}
=(\tilde k_a^*/2A_a^*)\langle{\bar K^0|H_w(-,-)|D^0}\rangle, \\
\langle{\pi^+|H_w(-,-)|\hat F^{*+}}\rangle = 0,
\end{array} 
\right.
\end{equation}
\begin{equation}
\left\{ 
\begin{array}{l}
\langle{\hat{\delta}^{s*+}|H_w(-,0)|D^+}\rangle 
=\sqrt{2}\langle{\hat{\delta}^{s*0}|H_w(-,0)|D^0 }\rangle \\
=\langle{\hat{\sigma}^{*}|H_w(-,0)|D^0 }\rangle            
=-\sqrt{2}\langle{\hat{\sigma}^{s*}|H_w(-,0)|D^0 }\rangle 
= (k_a^*/2A_a^*)\langle{\bar K^0|H_w(-,0)|D^0}\rangle, \\
-\sqrt{2}\langle{\pi^+|H_w(-,0)|\hat D^{*+}}\rangle 
=-{2}\langle{\pi^0|H_w(-,0)|\hat D^{*0}}\rangle \\
=\langle{K^+|H_w(-,0)|\hat F_I^{*+}}\rangle 
=\langle{K^0|H_w(-,0)|\hat F_I^{*0}}\rangle 
=(\tilde k_a^*/\sqrt{2}A_a^*)\langle{\bar K^0|H_w(-,0)|D^0}\rangle, \\ 
\langle{K^+|H_w(-,0)|\hat F^{*+}}\rangle = 0,
\end{array} 
\right.
\end{equation}
where $A_a^*$ is the invariant matrix element of axial charge defined 
by $A_a^*=-{1\over 2}\langle{\hat\kappa^{*+}|A_{\pi^+}|K^0}\rangle$. 

\hfil\break
(iii) Constraints on asymptotic matrix elements of $H_w$ between the
ground-state-meson and $(qq)(\bar q\bar q)$ meson states:
\begin{equation}
\left\{ 
\begin{array}{l}
\sqrt{{3\over 2}}\langle{E_{\pi\bar K}^{*+}|H_w(-,-)|D^+}\rangle 
=({3\over \sqrt{2}})\langle{E_{\pi\bar K}^{*0}|H_w(-,-)|D^0}\rangle \\ 
=3\langle{C_{\bar K}^{*0}|H_w(-,-)|D^0}\rangle               
=\sqrt{3}\langle{C_{\pi}^{s*+}|H_w(-,-)|D_s^+}\rangle         \\
\hspace{8cm}
= (k_s^*/ A_s^*)\langle{\bar K^0|H_w(-,-)|D^0}\rangle, \\
-\sqrt{{3\over 2}}\langle{\pi^+|H_w(-,-)|E_{\pi F}^{*+}}\rangle 
=\sqrt{{3\over 2}}\langle{\pi^0|H_w(-,-)|E_{\pi F}^{*0}}\rangle \\
= -({3\over\sqrt{2}})\langle{\bar K^0|H_w(-,-)|E_{\pi D}^{*0}}\rangle 
= \sqrt{3\over2}\langle{K^-|H_w(-,-)|E_{\pi D}^{*-}}\rangle \\
= 3\langle{\bar K^0|H_w(-,-)|C_{D}^{*0}}\rangle 
= -\sqrt{3}\langle{\bar K^0|H_w(-,-)|C_{D}^{s*0}}\rangle 
=\sqrt{3\over 2}\langle{K^+|H_w(-,-)|E_{KF}^{*+}}\rangle       \\
\hspace{8cm}
= (\tilde k_s^*/ A_s^*)\langle{\bar K^0|H_w(-,-)|D^0}\rangle, \\
\langle{\pi^+|H_w(-,-)|C_{F}^{*+}}\rangle = 0.
\end{array} 
\right.
\end{equation}
\begin{equation}
\left\{ 
\begin{array}{l}
\sqrt{3}\langle{E_{\pi\pi}^{*+}|H_w(-,0)|D^+}\rangle 
=({3\over \sqrt{2}})\langle{E_{\pi\pi}^{*0}|H_w(-,0)|D^0}\rangle     \\
=\sqrt{3}\langle{C_{\pi}^{s*+}|H_w(-,0)|D^+}\rangle       
=\sqrt{6}\langle{C_{\pi}^{s*0}|H_w(-,0)|D^0}\rangle              \\
=-\sqrt{3}\langle{C_{\pi}^{*+}|H_w(-,0)|D^+}\rangle         
=-3\langle{C^{*}|H_w(-,0)|D^0}\rangle          
=\sqrt{6}\langle{C^{s*}|H_w(-,0)|D^0}\rangle                    \\
=-({3\over\sqrt{2}})\langle{E_{\pi K}^{*+}|H_w(-,0)|D_s^+}\rangle 
=-3\langle{C_{K}^{*+}|H_w(-,0)|D_s^+}\rangle                  \\
\hspace{8cm}
= (k_s^*/A_s^*)\langle{\pi^+|H_w(-,0)|D^+}\rangle, \\
({3\over2\sqrt{2}})\langle{\pi^+|H_w(-,0)|E_{\pi D}^{*+}}\rangle 
=3\langle{\pi^0|H_w(-,0)|E_{\pi D}^{*0}}\rangle               \\
=-3\langle{\pi^+|H_w(-,0)|C_D^{*+}}\rangle 
=-3\sqrt{2}\langle{\pi^0|H_w(-,0)|C_D^{*0}}\rangle        \\   
\hspace{8cm}
= (\tilde k_s^*/A_s^*)\langle{\pi^+|H_w(-,0)|D^+}\rangle, \\
\langle{C_\pi^{*0}|H_w(-,0)|D^0}\rangle 
=\langle{K^+|H_w(-,0)|E_{\pi F}^{*+}}\rangle = 0, 
\end{array} 
\right.
\end{equation}
where $A_s^*$ is the invariant matrix element of axial charge defined 
by $A_s^*=\langle{C_K^{*+}|A_{\pi^+}|K^0}\rangle$. 

In the above, $k_a^*$ and $\tilde k_a^*$ ($k_s^*$ and $\tilde k_s^*$) 
are not generally equal to each other. However, use of the
commutation relation, 
$[[H_w(-,-),V_{D_s^-}],V_{D_s^-}] = [[H_w(-,-),V_{\pi^-}],V_{\pi^-}]$, 
with asymptotic $SU_f(4)$ symmetry (or an $SU_f(4)$ extension of the
nonet symmetry in the flavor $SU_f(3)$ with respect to asymptotic 
matrix elements of charges) leads to $k_a^* = \tilde k_a^*$ (and 
$k_s^* = \tilde k_s^*$). 

Matrix elements of $H_w(-,-)$ and $H_w(-,0)$ can be related to each
other, for example, as  
\begin{equation}
\langle{\pi^+|H_w(-,-)|D_s^+}\rangle 
= {V_{cs} \over V_{cd}}\langle{\pi^+|H_w(-,0)|D^+}\rangle,
\quad {\rm etc.}
\end{equation}
by using the commutation relations,
$\bigl[O_\pm(-,-),\,V_{K^0}\bigr] = O_\pm(-,0)$ 
and
$\bigl[O_\pm(-,0),\,V_{\bar K^0}\bigr] = 2O_\pm(-,-)$, 
where $O_\pm(-,-)$ and $O_\pm(-,0)$ are four quark operators in 
$H_w(-,-)$ and $H_w(-,0)$, respectively, and contributions of the QCD
induced penguin term have been neglected. 

\section{Non-factorizable amplitudes}

We here list hard pseudo-scalar-meson amplitudes as the
non-factorizable ones which include the ETC term describing continuum 
contributions and the surface term containing pole contributions of 
the ground-state $\{q\bar q\}_0$, scalar $[qq][\bar q\bar q]$ and 
$(qq)(\bar q\bar q)$ mesons and a glue-ball. They are revised from 
the ones given in Ref.\cite{TBD} in which the amplitudes involved 
some misprints and the insufficient parameterization of phases of the 
ETC terms.
\hfil\break
\noindent{(i) Cabibbo-angle-favored decays:}
\begin{eqnarray}
&&M_{non-f}(D^+\rightarrow \pi^+\bar K^0)          
\simeq -{i\over \sqrt{2}f_\pi}\langle{\pi^+|H_w|D^+_s}\rangle  
\nonumber\\
&&\times
\Biggl\{\Bigl(1-{f_\pi \over f_K}\Bigr)e^{i\delta_3}      
+\Bigl[\Bigl({m_D^2-m_K^2 \over m_{D^*}^2-m_K^2}\Bigr)
-\Bigl({m_D^2-m_\pi^2 \over m_{D_s^*}^2-m_\pi^2}\Bigr)
\Bigl({f_\pi \over f_K}\Bigr)\Bigr]k_0  \nonumber\\
&&{\hspace{2cm}}\qquad\,
+\Bigl[\Bigl({m_D^2-m_K^2 \over m_{\hat D^*}^2-m_K^2}\Bigr)
-\Bigl({m_D^2-m_\pi^2 \over m_{\hat F_I^*}^2-m_\pi^2}\Bigr)
\Bigl({f_\pi \over f_K}\Bigr)\Bigr]k_a^*            \nonumber\\
&&{\hspace{2.9cm}}
+\Bigl[2\Bigl({m_D^2-m_K^2 \over m_D^2-m_{E_{\pi K}^*}^2}\Bigr)
+2\Bigl({m_D^2-m_\pi^2 \over m_D^2-m_{E_{\pi K}^*}^2}\Bigr)
\Bigl({f_\pi \over f_K}\Bigr)          \nonumber\\            
&&{\hspace{3.5cm}}
-\Bigl({m_D^2-m_K^2 \over m_{E_{\pi D}^*}^2-m_K^2}\Bigr)
-\Bigl({m_D^2-m_\pi^2 \over m_{E_{\pi F}^*}^2-m_\pi^2}\Bigr)
\Bigl({f_\pi \over f_K}\Bigr)\Bigr]k_s^*
\Biggr\},
\end{eqnarray}
\begin{eqnarray}
&&M_{non-f}(D^0 \rightarrow \pi^+K^-)                  \nonumber\\
&&\simeq {i\over \sqrt{2}f_\pi}\langle{\pi^+|H_w|D_s^+}\rangle  
\Biggl\{
-{1 \over 3}\Bigl(1-{f_\pi \over f_K}\Bigr)e^{i\delta_3(\pi\bar K)}
+{1 \over 3}\Bigl(4-{f_\pi \over f_K}\Bigr)e^{i\delta_1(\pi\bar K)}   
                                                         \nonumber\\
&&{\hspace{4cm}}+\Bigl[\Bigl({m_D^2-m_K^2 \over m_D^2-m_{K^*}^2}\Bigr)
-\Bigl({m_D^2-m_\pi^2 \over m_D^2-m_{K^*}^2}\Bigr)
\Bigl({f_\pi \over f_K}\Bigr)
+\Bigl({m_D^2-m_\pi^2 \over m_{D_s^*}^2-m_\pi^2}\Bigr)
\Bigl({f_\pi \over f_K}\Bigr)\Bigr]k_0     \nonumber\\
&&{\hspace{4cm}}
+\Bigl[\Bigl({m_D^2-m_K^2 \over m_D^2-m_{\hat\kappa^*}^2}\Bigr)
+\Bigl({m_D^2-m_\pi^2 \over m_D^2-m_{\hat\kappa^*}^2}\Bigr)
\Bigl({f_\pi \over f_K}\Bigr)
-\Bigl({m_D^2-m_\pi^2 \over m_{\hat F_I^*}^2-m_\pi^2}\Bigr)
\Bigl({f_\pi \over f_K}\Bigr)
\Bigr]k_a^*             \nonumber\\
&&{\hspace{4cm}}
-\Bigl[\Bigl({m_D^2-m_K^2 \over m_D^2-m_{E_{\pi K}^*}^2}\Bigr)
+\Bigl({m_D^2-m_\pi^2 \over m_D^2-m_{E_{\pi K}^*}^2}\Bigr)
\Bigl({f_\pi \over f_K}\Bigr)                       \nonumber\\
&&{\hspace{4cm}}\qquad
-2\Bigl({m_D^2-m_K^2 \over m_{E_{\pi D}^*}^2-m_K^2}\Bigr)
+\Bigl({m_D^2-m_\pi^2 \over m_{E_{\pi F}^*}^2-m_\pi^2}\Bigr)
\Bigl({f_\pi \over f_K}\Bigr)\Bigr]k_s^*
\Biggr\},
\end{eqnarray}
\begin{eqnarray}
&&M_{non-f}(D^0 \rightarrow \pi^0\bar K^0)                  \nonumber\\
&&\simeq -{i\over \sqrt{2}f_\pi}\langle{\pi^+|H_w|D_s^+}\rangle  
\sqrt{1\over 2}
\Biggl\{
-{\sqrt{2} \over 3}\Bigl(1-{f_\pi \over f_K}\Bigr)
                                               e^{i\delta_3(\pi\bar K)}
-{\sqrt{2} \over 3}\Bigl(2-{1 \over 2}{f_\pi \over f_K}\Bigr)
                                               e^{i\delta_1(\pi\bar K)}        
                                                           \nonumber\\
&&{\hspace{4cm}}\quad
+\Bigl[\Bigl({m_D^2-m_K^2 \over m_D^2-m_{K^*}^2}\Bigr)
-\Bigl({m_D^2-m_\pi^2 \over m_D^2-m_{K^*}^2}\Bigr)
\Bigl({f_\pi \over f_K}\Bigr)
+\Bigl({m_D^2-m_K^2 \over m_{D^*}^2-m_K^2}\Bigr)\Bigr]k_0 \nonumber\\
&&{\hspace{4cm}}\quad
+\Bigl[\Bigl({m_D^2-m_K^2 \over m_D^2-m_{\hat\kappa^*}^2}\Bigr)
+\Bigl({m_D^2-m_\pi^2 \over m_D^2-m_{\hat\kappa^*}^2}\Bigr)
                          \Bigl({f_\pi \over f_K}\Bigr)  \nonumber\\
&&{\hspace{5cm}}
+\Bigl({m_D^2-m_K^2 \over m_{\hat D^*}^2-m_K^2}\Bigr)
-2\Bigl({m_D^2-m_\pi^2 \over m_{\hat F_I^*}^2-m_\pi^2}\Bigr)
\Bigl({f_\pi \over f_K}\Bigr)\Bigr]k_a^*             \nonumber\\
&&{\hspace{4cm}}\quad
+\Bigl[\Bigl({m_D^2-m_K^2 \over m_D^2-m_{E_{\pi K}^*}^2}\Bigr)
+\Bigl({m_D^2-m_\pi^2 \over m_D^2-m_{E_{\pi K}^*}^2}\Bigr)
                       \Bigl({f_\pi \over f_K}\Bigr) \nonumber\\
&&{\hspace{5cm}}
+\Bigl({m_D^2-m_K^2 \over m_{E_{\pi D}^*}^2-m_K^2}\Bigr)
-2\Bigl({m_D^2-m_\pi^2 \over m_{E_{\pi F}^*}^2-m_\pi^2}\Bigr)
\Bigl({f_\pi \over f_K}\Bigr)\Bigr]k_s^*
\Biggr\},
\end{eqnarray}
\begin{eqnarray}
&&M_{non-f}(D_s^+ \rightarrow K^+\bar K^0)    \nonumber\\
&&\simeq -{i \over \sqrt{2}f_K}\langle{\pi^+|H_w|D_s^+}\rangle
\Biggl\{
e^{i\delta_2(K\bar K)}+\Bigl(
{m_{D_s}^2-m_K^2 \over m_{D^*}^2-m_K^2}\Bigr)k_0       \nonumber\\
&&{\hspace{4cm}}\quad 
+\Bigl[
2\Bigl({m_{D_s}^2-m_K^2 \over m_{D_s}^2-m_{\hat\delta^{s*}}^2}\Bigr)
-\Bigl({m_{D_s}^2-m_K^2 \over m_{\hat D^{s*}}-m_K^2}\Bigr)\Bigr]k_a^* 
\nonumber\\
&&{\hspace{4cm}}\quad 
+2\Bigl[\Bigl({m_{D_s}^2-m_K^2 \over m_{D_s}^2-m_{C_\pi^s}^2}\Bigr)
-\Bigl({m_{D_s}^2-m_K^2 \over m_{E_{KF}}^2-m_K^2}\Bigr)\Bigr]
                                                      k_s^*\Biggr\}, 
\end{eqnarray}
(ii) Cabibbo-angle suppressed decays:
\begin{eqnarray}
&&M_{non-f}(D^0 \rightarrow \pi^+\pi^-)    \nonumber\\
&&\simeq {i\over \sqrt{2}f_\pi}\langle{\pi^+|H_w|D^+}\rangle
\Biggl\{
e^{i\delta_0(\pi\pi)}
+ \Bigl({m_D^2-m_\pi^2 \over m_{D^*}^2-m_\pi^2}\Bigr)k_0 
\nonumber\\
&&{\hspace{4cm}} 
+\Bigl[2\Bigl({m_D^2-m_\pi^2 \over m_D^2-m_{\hat\sigma^*}^2}\Bigr)
-\Bigl({m_D^2-m_\pi^2 \over m_{\hat D^*}^2-m_\pi^2}\Bigr)\Bigr]k_a^*
                                                         \nonumber\\
&&{\hspace{4cm}}
-\Bigl[2\Bigl({m_D^2-m_\pi^2 \over m_D^2-m_{E_{\pi\pi}}^2}\Bigr)
-\Bigl(
{m_D^2-m_\pi^2 \over m_{E_{\pi D}}^2-m_\pi^2}\Bigr)\Bigr]k_s^*
-{2 \over Z}\Bigl({m_D^2-m_\pi^2 \over m_D^2-m_{S^*}^2}\Bigr)k_g
\Biggr\},
\end{eqnarray}
\begin{eqnarray}
&&M_{non-f}(D^0 \rightarrow \pi^0\pi^0)    \nonumber\\
&&\simeq {i\over \sqrt{2}f_\pi}\langle{\pi^+|H_w|D^+}\rangle
\sqrt{1 \over 2}\Biggl\{
e^{i\delta_0(\pi\pi)}
+ \Bigl({m_D^2-m_\pi^2 \over m_{D^*}^2-m_\pi^2}\Bigr)k_0 \nonumber\\
&&\hspace{4cm}\qquad 
+\Bigl[2\Bigl({m_D^2-m_\pi^2 \over m_D^2-m_{\hat\sigma^*}^2}\Bigr)
-\Bigl({m_D^2-m_\pi^2 \over m_{\hat D^*}^2-m_\pi^2}
                                     \Bigr)\Bigr]k_a^* \nonumber\\
&&{\hspace{4cm}}
+\Bigl[2\Bigl({m_D^2-m_\pi^2 \over m_D^2-m_{E_{\pi\pi}}^2}\Bigr)
-\Bigl(
{m_D^2-m_\pi^2 \over m_{E_{\pi D}}^2-m_\pi^2}\Bigr)\Bigr]k_s^*
-{2 \over Z}\Bigl({m_D^2-m_\pi^2 \over m_D^2-m_{S^*}^2}\Bigr)k_g
\Biggr\},
\end{eqnarray}
\begin{eqnarray}
&&M_{non-f}(D^+ \rightarrow \pi^+\pi^0)   
\simeq {i\over \sqrt{2}f_\pi}\langle{\pi^+|H_w|D^+}\rangle
\Biggl\{
\Bigl[2\Bigl({m_D^2-m_\pi^2 \over m_D^2-m_{E_{\pi\pi}}^2}\Bigr)
-\Bigl(
{m_D^2-m_\pi^2 \over m_{E_{\pi D}}^2-m_\pi^2}\Bigr)\Bigr]\sqrt{2}k_s^*
\Biggr\},
\end{eqnarray}
\begin{eqnarray}
&&M_{non-f}(D^0 \rightarrow K^+K^-)    \nonumber\\
&&\simeq -{i\over \sqrt{2}f_K}\langle{\pi^+|H_w|D^+}\rangle
\Biggl\{
e^{i\delta_2(K\bar K)}
+ \Bigl({m_D^2-m_K^2 \over m_{D_s^*}^2-m_K^2}\Bigr)k_0  \nonumber\\
&&{\hspace{4cm}}\quad
+\Bigl[2\Bigl({m_D^2-m_K^2 \over m_D^2-m_{\hat\sigma^{s*}}^2}\Bigr)
-{m_D^2-m_K^2 \over m_{\hat F_I^*}^2-m_K^2}\Bigr]k_a^* 
\nonumber\\
&&{\hspace{4cm}}\quad 
-\Bigl[2\Bigl({m_D^2-m_K^2 \over m_D^2-m_{C^{s*}}^2}\Bigr)
-2\Bigl({m_D^2-m_K^2 \over m_{E_{\bar K D}^*}^2-m_K^2}\Bigr)
+\Bigl({m_D^2-m_K^2 \over m_{C_F^*}^2-m_K^2}\Bigr)\Bigr]k_s^*
\nonumber\\
&&{\hspace{10cm}}\qquad
+2\Bigl({m_D^2-m_\pi^2 \over m_D^2-m_{S^*}^2}\Bigr)k_g
\Biggr\},
\end{eqnarray}
\begin{eqnarray}
&&M_{non-f}(D^0 \rightarrow K^0\bar K^0)    \nonumber\\
&&\simeq {i\over \sqrt{2}f_K}\langle{\pi^+|H_w|D^+}\rangle
\Biggl\{\Bigl[\Bigl(
{m_D^2-m_K^2 \over m_{E_{\bar KD}}^2-m_K^2}
\Bigr)
-2\Bigl({m_D^2-m_K^2 \over m_{E_{\pi F}}^2-m_K^2}\Bigr)
\Bigr]k_s^*
-2\Bigl({m_D^2-m_K^2 \over m_D^2-m_{S^*}^2}\Bigr)k_g\Biggr\}, 
\end{eqnarray}
\begin{eqnarray}
&&M_{non-f}(D^+ \rightarrow K^+\bar K^0)    \nonumber\\
&&\simeq -{i\over \sqrt{2}f_K}\langle{\pi^+|H_w|D^+}\rangle
\Biggl\{
e^{i\delta_2(K\bar K)}                      
+ \Bigl({m_D^2-m_K^2 \over m_{D_s^*}^2-m_K^2}\Bigr)k_0
                                                        \nonumber\\
&&\hspace{4cm}\quad
+\Bigl[2\Bigl({m_D^2-m_K^2 \over m_D^2-m_{\hat\delta^{s*}}^2}\Bigr)
-2\Bigl({m_D^2-m_K^2 \over m_{\hat F_0^*}^2-m_K^2}\Bigr)
+\Bigl({m_D^2-m_K^2 \over m_{\hat F_I^*}^2-m_K^2}\Bigr)
\Bigr]k_a^* 
\nonumber\\
&&\hspace{4cm}\quad
-\Bigl[2\Bigl({m_D^2-m_K^2 \over m_D^2-m_{C_\pi^{s*}}^2}\Bigr)
-\Bigl({m_D^2-m_K^2 \over m_{E_{\bar KD}^*}^2-m_K^2}\Bigr)
+\Bigl({m_D^2-m_K^2 \over m_{C_F^*}^2-m_K^2}\Bigr)\Bigr]k_s^*
\Biggr\},
\end{eqnarray}
\begin{eqnarray}
&&M_{non-f}(D_s^+ \rightarrow \pi^+K^0)     \nonumber\\
&&\simeq  {i\over \sqrt{2}f_\pi}\langle{\pi^+|H_w|D^+_s}\rangle  
\Biggl\{\Bigl[2\Bigl({f_\pi \over f_K}\Bigr)-1\Bigr]
                            e^{i\delta_1(\pi K)}    \nonumber\\
&&\hspace{4cm}
-\Bigl[\Bigl({m_{D_s}^2-m_K^2 \over m_{D_s}^2-m_{K^*}^2}\Bigr)
-\Bigl({m_{D_s}^2-m_\pi^2 \over m_{D_s}^2-m_{K^*}^2}\Bigr)
\Bigl({f_\pi \over f_K}\Bigr)
-\Bigl({m_{D_s}^2-m_\pi^2 \over m_{D^*}^2-m_\pi^2}\Bigr)
\Bigl({f_\pi \over f_K}\Bigr)\Bigr]k_0            \nonumber\\
&&\hspace{4cm}
+\Bigl[
\Bigl({m_{D_s}^2-m_K^2 \over m_{D_s}^2-m_{\hat\kappa^*}^2}\Bigr)
+\Bigl({m_{D_s}^2-m_\pi^2 \over m_{D_s}^2-m_{\hat\kappa^*}^2}\Bigr)
\Bigl({f_\pi \over f_K}\Bigr) \nonumber\\
&&\hspace{8cm}
-2\Bigl({m_{D_s}^2-m_K^2 \over m_{\hat F_I^*}^2-m_K^2}\Bigr)
+\Bigl({m_{D_s}^2-m_\pi^2 \over m_{\hat F^*}^2-m_\pi^2}\Bigr)
\Bigl({f_\pi \over f_K}\Bigr)\Bigr]k_a^*            \nonumber\\
&&\hspace{4cm}
-\Bigl[\Bigl({m_{D_s}^2-m_K^2 \over m_{D_s}^2-m_{E_{\pi K}^*}^2}\Bigr)
+\Bigl({m_{D_s}^2-m_\pi^2 \over m_{D_s}^2-m_{E_{\pi K}^*}^2}\Bigr)
\Bigl({f_\pi \over f_K}\Bigr) \nonumber\\
&&\hspace{7cm}
-2\Bigl({m_{D_s}^2-m_K^2 \over m_{E_{\pi F}^*}^2-m_K^2}\Bigr)
+\Bigl({m_{D_s}^2-m_\pi^2 \over m_{C_D^{s*}}^2-m_\pi^2}\Bigr)
\Bigl({f_\pi \over f_K}\Bigr)\Bigr]k_s^*
\Biggr\},
\end{eqnarray}
\begin{eqnarray}
&&M_{non-f}(D_s^+ \rightarrow \pi^0K^+)     \nonumber\\
&&\simeq {i\over \sqrt{2}f_\pi}\langle{\pi^+|H_w|D^+_s}\rangle  
\sqrt{1\over 2}\Biggl\{\Bigl[2\Bigl({f_\pi \over f_K}\Bigr)-1\Bigr]
                            e^{i\delta_1(\pi K)}    \nonumber\\
&&\hspace{4.5cm}\,\,
-\Bigl[\Bigl({m_{D_s}^2-m_K^2 \over m_{D_s}^2-m_{K^*}^2}\Bigr)
-\Bigl({m_{D_s}^2-m_\pi^2 \over m_{D_s}^2-m_{K^*}^2}\Bigr)
\Bigl({f_\pi \over f_K}\Bigr)                       \nonumber\\
&&\hspace{7.8cm}
-\,\,\Bigl({m_{D_s}^2-m_\pi^2 \over m_{D^*}^2-m_\pi^2}\Bigr)
\Bigl({f_\pi \over f_K}\Bigr)\Bigr]k_0  \nonumber\\
&& \hspace{4.5cm}\,\,
+\Bigl[
\Bigl({m_{D_s}^2-m_K^2 \over m_{D_s}^2-m_{\hat\kappa^*}^2}\Bigr)
+\Bigl({m_{D_s}^2-m_\pi^2 \over m_{D_s}^2-m_{\hat\kappa^*}^2}\Bigr)
\Bigl({f_\pi \over f_K}\Bigr)   \nonumber\\
&&\hspace{7.0cm}
-2\Bigl({m_{D_s}^2-m_K^2 \over m_{\hat F_I^*}^2-m_K^2}\Bigr)
+\Bigl({m_{D_s}^2-m_\pi^2 \over m_{\hat F^*}^2-m_\pi^2}\Bigr)
\Bigl({f_\pi \over f_K}\Bigr)\Bigr]k_a^*            \nonumber\\
&&\hspace{4.5cm}\,\,
+\Bigl[\Bigl({m_{D_s}^2-m_K^2 \over m_{D_s}^2-m_{E_{\pi K}^*}^2}\Bigr)
+\Bigl({m_{D_s}^2-m_\pi^2 \over m_{D_s}^2-m_{E_{\pi K}^*}^2}\Bigr)
\Bigl({f_\pi \over f_K}\Bigr)   \nonumber\\
&&\hspace{9.5cm}\quad
-\Bigl({m_{D_s}^2-m_\pi^2 \over m_{C_D^{s*}}^2-m_\pi^2}\Bigr)
\Bigl({f_\pi \over f_K}\Bigr)\Bigr]k_s^*
\Biggr\},
\end{eqnarray}



\begin{references}
\bibitem{phase} In Ref.\cite{TBD}, we neglected  possible phase 
differences among different isospin eigen amplitudes since
contributions of ETC terms for decays into final states with exotic 
quantum number like $I={3\over 2}$ are very small (vanishing in the 
$SU_f(3)$ symmetry) because of the selection rule from 
Eq.(\ref{eq:SUM-G}). It has been improved in this article.
\bibitem{f_0} In Ref.\cite{TBD}, we took $|Z| \rightarrow\infty$ 
since we assigned $f_0(1770)$ to $S^*$. It had been reported to decay 
dominantly into $K\bar K$ (but not $\pi\pi$) final states. However, 
it is now called as $f_J(1710)$ and can decay into $\pi\pi$ final 
states.
\end{references}
\end{document}